# Controlling the Oscillator Frequency Synchronization by a Low-Frequency Drive

Anastasiia Y. Nimets[1,2], Klaus Schuenemann[1], Dmytro M. Vavriv[2]

[1]*Institute of High-Frequency Technology, Hamburg University of Technology*
*22 Denickestrasse, 21073 Hamburg, Germany*
anastasiia.nimets@tuhh.de
schuenemann@tuhh.de

[2]*Department of Microwave Electronics, Institute of Radio Astronomy of NAS of Ukraine*
*4 Chervonopraporna Str., 61002 Kharkov, Ukraine*
vavriv@rian.kharkov.ua

*Abstract*— The dynamics of an oscillator driven by both low- and high- frequency external signals are studied. It is shown that both two- and three-frequency resonances arise due to a nonlinear interaction of these harmonic forces. Conditions which must be met for oscillator synchronization under these resonances are estimated analytically by considering the Van der Pol oscillator with modulated natural frequency as mathematical model. It is demonstrated that due to the low-frequency modulation, additional synchronization regions arise in the control parameter space. Feasibility of the theoretical findings is confirmed in experiments with a Hartley-type oscillator.

## I. Introduction

Synchronization of oscillators by a harmonic external forcing is widely used in many applications. This phenomenon is well understood due to an intensive theoretical study starting from the classical works in this field [1], [2]. It is assumed that a self-sustained oscillator is synchronized, if the following condition is met

$$\frac{\omega}{\omega_e} = \frac{n}{m}, \qquad (1)$$

where $\omega$ is the synchronized frequency, $\omega_e$ is the external synchronizing frequency, $n$ and $m$ are real integers.

Later on, the theory of synchronization was generalized for multi-mode and distributed systems, for oscillators with chaotic behaviour, and diverse factors affecting the synchronization were studied as well [3] - [6].

In this paper, we consider the synchronization of a self-sustained oscillator, when an additional low-frequency signal is applied to the oscillator. This signal is presented as a low-frequency modulation of the natural frequency of the oscillator. A generalized Van der Pol oscillator is considered as the mathematical model to study this problem. We show theoretically that the interaction of low- to high-frequency oscillations leads to an arising of additional ranges of synchronization. We find analytically the location of these ranges in the parameter space. We also present an experimental confirmation of the theoretical results by studying the synchronization of a single-transistor Hartley-type oscillator with modulated frequency of oscillation.

The paper is organized as follows. In Section II, the generalised Van der Pol oscillator with a high-frequency driving force and a low-frequency modulation of the natural frequency is presented, and the corresponding averaged equations are obtained. Section III describes resonances that can occur in the oscillator and the corresponding synchronization ranges are determined analytically by using the secondary averaging technique. In Section IV, the experimental data obtained with the Hartley-type oscillator are considered and compared with the theoretical results. Section V summarizes the main results of the paper.

## II. Mathematical Model

To study theoretically the interaction of low- to high-frequency oscillations in self-sustained oscillators, we consider a general Van der Pol oscillator

$$\frac{d^2 x}{dt^2} + \omega(t)^2 x = 2\omega_0(\alpha - a_2 x^2)\frac{dx}{dt} + 2\omega_0^2 F_e \sin\omega_e t. \qquad (2)$$

Here $x$ is the generalized coordinate, $\alpha > 0$ is the increment of the oscillation growth, $a_2 > 0$ is the coefficient, which determines the amplitude dependent dissipation in the system, $F_e$ is the amplitude of the high-frequency external forcing with the corresponding frequency $\omega_e$. The natural frequency of the oscillator $\omega(t)$ varies harmonically in time

$$\omega(t) = \omega_0(1 + m\cos\Omega t), \qquad (3)$$

where $\omega_0$ is a mean value of the natural frequency, $m \ll 1$ is the modulation coefficient, $\Omega$ is the frequency of modulation.

We consider the case when

$$\omega_e \approx \omega_0, \qquad (4)$$

what means that just the ($n$:$m$=1:1)-mode of the synchronization (see (1)) can occur in the oscillator, when there is no low-frequency driving.

Considering the terms with $m$, $\alpha$, $a_2$, and $F_e$ in (2) as small perturbation, and assuming that

$$\alpha\omega_0 \ll \Omega \ll \omega_0, \quad (5)$$

we can apply the averaging technique [7] to this equation, resulting in the following equations with respect to both the amplitude $A$ and the phase $\varphi$ of the high-frequency oscillation excited in the oscillator

$$\frac{dA}{d\tau} = \alpha A - a_2 A^3/4 - F_e \cos\varphi,$$
$$\frac{d\varphi}{d\tau} = \Delta + 2m \cdot \cos\theta\tau + (F_e/A)\sin\varphi, \quad (6)$$

where $\tau = \omega_0 \cdot t$, $\Delta = \frac{\omega_0 - \omega_e}{\omega_0}$, $\theta = \frac{\Omega}{\omega_0}$.

When the low-frequency driving is not applied ($m=0$), we have from (6) the well known set of equations describing the dynamics of the harmonically excited oscillator:

$$\frac{dA}{d\tau} = A(\alpha - a_2 A^2/4) - F_e \cos\varphi,$$
$$\frac{d\varphi}{d\tau} = \Delta + (F_e/A)\sin\varphi. \quad (7)$$

The solutions of (7) are already understood in detail [2]. In particular, it has been shown that provided $F_e$ is sufficiently small, the oscillator is synchronized at the frequency $\omega_e$, if the external frequency is in the following range

$$\omega_0(1 - F_e/A_0) \leq \omega_e \leq \omega_0(1 + F_e/A_0), \quad (8)$$

where $A_0 = 2(\alpha/a_2)^{1/2}$ is the amplitude of the oscillation of the free-running oscillator. Thus the bandwidth of the synchronization $\sigma_{s2}$ for this case is

$$\sigma_{s2} = 2\omega_0 F_e/A_0. \quad (9)$$

It should be noted that the synchronization here occurs due to the two-frequency resonance (4). In the next Section, we consider the synchronization conditions for the oscillator when both low- and high-frequency signals are applied.

### III. INTERACTION OF HIGH- AND LOW-FREQUENCY OSCILLATIONS

A synchronization of the oscillations is possible due to resonances that can occur in a particular oscillator. In the considered oscillator, according to the governing equations (6), the following resonances can take place: $\Delta = 0$ and $\Delta = \pm\theta_2$. In term of the dimensional frequencies, the first of these resonances is the two-frequency resonance (4), while the next ones are three-frequency resonances:

$$\omega_e - \omega_0 \approx \pm\Omega_2. \quad (10)$$

Other higher-order resonances are possible in (6), but we restrict our further considerations to the above resonances.

By using the secondary averaging technique, (6) can be simplified when considering the oscillator dynamics in the vicinity of each particular resonance - (4) or (10). In particular, it is easy to show that the resonance (4) is also described by the set of equations (7) even if the low-frequency forcing is applied. It means that such forcing does not change the synchronization conditions and that it does not perturb the range of synchronization (8), which corresponds to the case of a pure high-frequency harmonic forcing.

For the resonances (10), equations (6) can be simplified in the following way. We introduce a new amplitude $B(\tau)$ and a phase $\psi(\tau)$, which are related to $A(\tau)$ and $\varphi(\tau)$ by the following relations

$$A\cos(\varphi) = B\cos[\pm 2\theta\tau + \psi(t)] + F_e/\Delta,$$
$$A\sin(\varphi) = B\sin[\pm 2\theta\tau + \psi(t)].$$

Then we use the secondary averaging technique arriving at the following set of equations

$$\frac{dB}{d\tau} = (\alpha_1 - \frac{a_2 B^2}{4})B + \frac{F_e}{\Omega}m\cos\psi,$$
$$B\frac{d\psi}{d\tau} = (\Delta \mp \Omega)B - \frac{F_e}{\Omega}m\sin\psi, \quad (11)$$

where $\alpha_1 = \alpha - \frac{F_e^2}{2\Omega^2}a_2$. Provided the amplitude $F_e$ is small, we have $\alpha_1 \approx \alpha$.

Nontrivial solutions of (11) exist only if both the amplitude $F_e$ and the modulation coefficient $m$ are not zero. Considering steady states ($dB/dt=d\psi/dt=0$) of (11), we find that the oscillator frequency is locked to the external frequency $\omega_e$ in the following frequency ranges

$$\omega_0 \pm \Omega - \frac{\omega_0^2 m F_e}{\Omega A_0} \leq \omega_e \leq \omega_0 \pm \Omega + \frac{\omega_0^2 m F_e}{\Omega A_0}. \quad (12)$$

Thus two additional synchronization ranges appear as the result of the three-frequency interaction (10). On the parameter plane in Fig. 1 with coordinates $\omega_e$ and $F_e$, these ranges, noted hereinafter as $SA_{3+}$ and $SA_{3-}$, are located symmetrically with respect to the synchronization range (8), noted as $SA_2$. According to (12), the bandwidth of the synchronization $\sigma_{s3}$ for each of the three-frequency resonances is

$$\sigma_{s3} = \frac{2m\omega_0^2 F_e}{\Omega A_0}. \quad (13)$$

Thus the bandwidth can be increased by increasing the modulation coefficient $m$ and/or the amplitude $F_e$.

The value of $\sigma_{s3}$ is also growing up when decreasing the modulation frequency, but it should be taken into account that the $\Omega$-value is constrained by the condition (5).

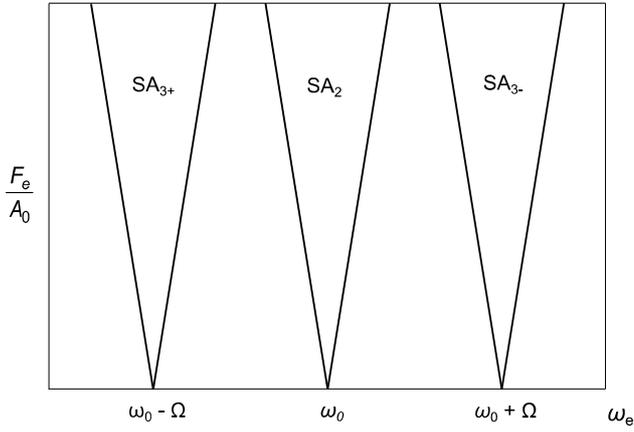

Fig. 1 Synchronization ranges, which arise due to two-frequency (SA$_2$) and three-frequency resonances (SA$_{3+}$ and SA$_{3-}$)

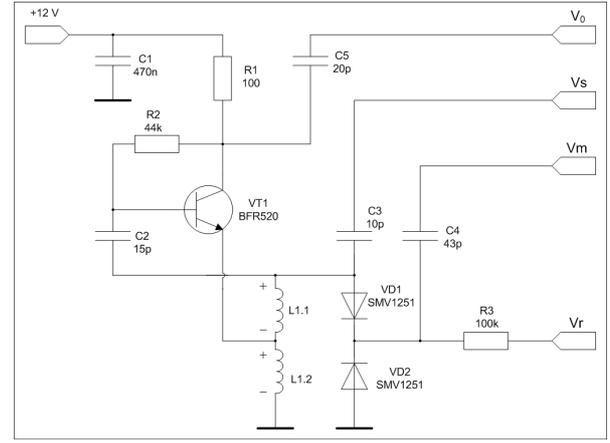

Fig. 2 Circuit diagram of the generator

Comparing $\sigma_{s3}$ with that value for the two-frequency resonance (9) we find

$$\frac{\sigma_{s3}}{\sigma_{s2}} = \frac{\omega_0 m}{\Omega}. \quad (14)$$

This ratio indicates that the bandwidth $\sigma_{s3}$ can be larger than $\sigma_{s2}$ even if a weak low-frequency modulation (small values of $m$) is applied to the oscillator. Hence such modulation enables to provide synchronization of an oscillator in a wider frequency domain. The location of the additional ranges of synchronization is determined by both the frequency and the amplitude of the modulation signal what is important for practical applications.

The analysis of the steady-state amplitudes for the synchronized oscillations, which depend on the resonances considered, shows that these amplitudes are close (with an accuracy of the forcing amplitude $F_e$) to the amplitude of the self-running oscillator $A_0$.

## IV. EXPERIMENT

The described theoretical findings can be easily verified experimentally by using simple oscillator circuits. We have used the Hartley-type oscillator shown in Fig. 2. The tuned circuit consists of two varactor diodes SMV1251 and two inductors in series. We have applied a reverse varactor voltage of $V_r$=1.5 V, what resulted in a frequency $f_0$ of the free-running oscillator of 107.5 MHz. A high-frequency synchronizing signal $V_s$ and a low-frequency modulating signal $V_m$ from external generators are applied to the oscillator as shown in Fig. 2.

At first, we have determined the synchronization range, when just a high frequency signal $V_s$ is applied. This range, which corresponds to the 1:1 synchronization condition (1) and arises due to the two-frequency resonance (4), is shown in Fig. 3 on the parameter plane as synchronizing frequency $f_s$ and amplitude of the synchronizing signal.

After that, low frequency modulation signals with $m$=0.1 and frequencies of modulation $f_m$ of 7 MHz and 10 MHz respectively, have been applied.

The corresponding synchronization ranges are shown in Fig. 4 and Fig. 5. On each of the figures, there are three synchronization ranges SA$_2$, SA$_{3+}$, and SA$_{3-}$. The application of the modulation signal changes just slightly SA$_2$ in both dimensions and location, as follows from the theoretical results. The spectrum of the synchronized oscillations, shown in Fig. 6, is the same for each of the synchronized ranges, while the spectrum of the output non-synchronized oscillations outside these ranges is like that in Fig. 7.

The ranges SA$_{3+}$, and SA$_{3-}$, which are due to the three frequencies (10), have dimensions and locations in good agreement with the theoretical predictions (12) - (14). For example, from Fig. 5, it follows that the width of the synchronization ranges is approximately equal for all values of the synchronizing amplitude. This means that the ratio $\sigma_{s3}/\sigma_{s2} \approx 1$. From (14), we find that in order to have such ratio for the considered case, the modulation coefficient $m$ should be 0.1. Hence this value coincides with that used in the experiment. The same value for $m$ follows from (14) and from the experimental data in Fig. 4.

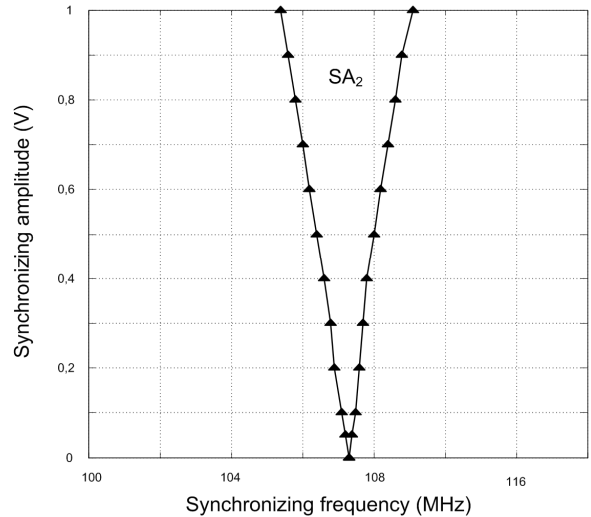

Fig. 3 Synchronization range of unmodulated oscillator

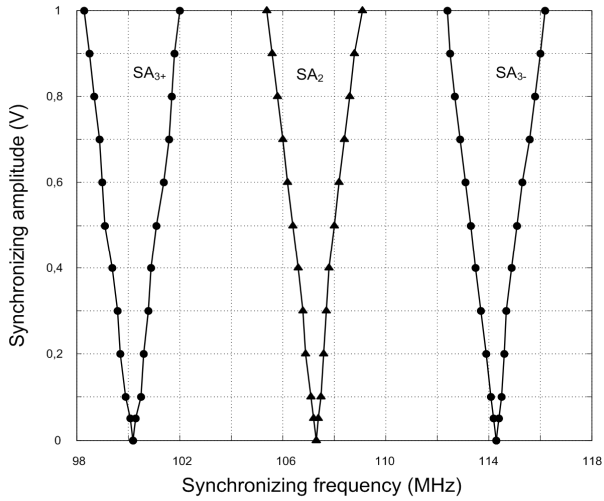

Fig.4 Synchronization ranges with the modulation frequency of 7 MHz

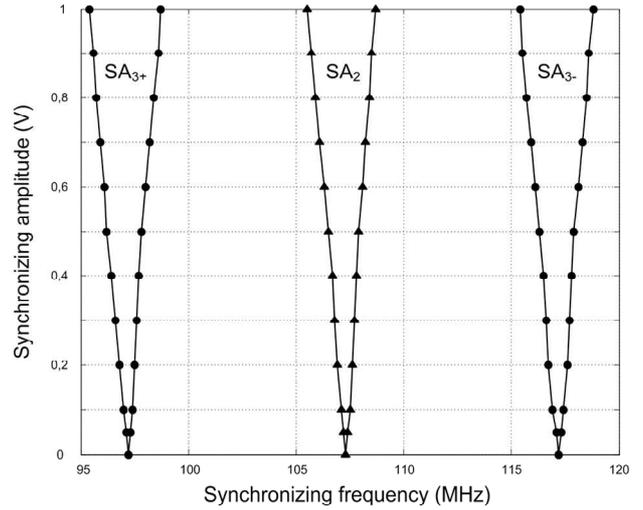

Fig. 5 Synchronization ranges with modulation frequency of 10 MHz

It can also be seen from Fig. 4 and Fig. 5 that the decrease of $f_m$ leads to increasing the width of $SA_{3+}$ and $SA_{3-}$ and decreasing their distance to the range $SA_2$, what is also predicted by the theoretical results.

## V. CONCLUSION

Thus the low-frequency modulation of the natural frequency of the oscillator enables to control effectively the synchronization ranges of the synchronized oscillator. In particular, additional ranges of synchronization arise due to such modulation, their location and dimensions are determined by both the amplitude and the frequency of the modulation. Analytical expressions are found to describe accurately synchronization conditions and ranges of synchronization. Experiments have well confirmed the theoretical results.

ACKNOWLEDGMENT

A. Y. N. would like to acknowledge DAAD for financial support.

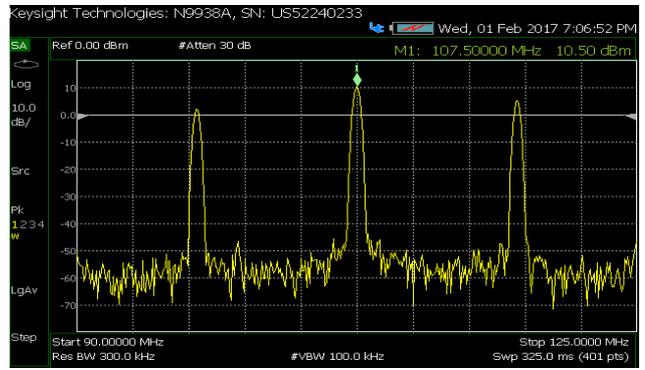

Fig. 6 Spectrum of synchronized oscillation, when the modulation of 10 MHz is applied

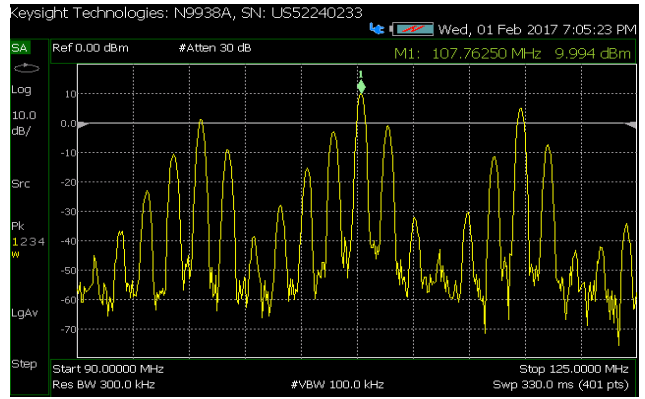

Fig. 7 Spectrum of non-synchronized oscillation, when the modulation of 10 MHz is applied